\documentclass[12pt]{article}
\usepackage{amssymb}
\usepackage{stmaryrd}
\title{Playful, streamlike computation}
\date{January 2002} 
\author{Pierre-Louis Curien (CNRS -- Universit\'e Paris VII)}

\begin{document}
\maketitle

\newtheorem{theorem}{Theorem}[section]
\newtheorem{definition}[theorem]{Definition}
\newtheorem{proposition}[theorem]{Proposition}
\newtheorem{exercice}[theorem]{Exercice}
\newtheorem{hardexercise}[theorem]{Exercise${}^*$}
\newtheorem{remark}[theorem]{Remark}
\newtheorem{example}[theorem]{Example}

\newenvironment{branch}{\left\{\begin{array}{l}}{\end{array}\right.}

\newcommand{\qedm}{\mbox{}\\[-2.5em] \mbox{}\hfill\rule{6pt}{6pt}} 
\newcommand{\Preuve}{\noindent {\sc Preuve}. }
\newcommand{\Preuvehint}{\noindent {\sc Preuve (indication)}. }
\newcommand{\Proofitem}[1]{\medskip \noindent $#1\;$}
\newcommand{\Proofitemf}[1]{ $#1\;$}

\newcommand{\guil}[1]{``#1"}
\newcommand{\lbd}{\lambda}
\newcommand{\mand}{\mbox{ and }}
\newcommand{\mor}{\mbox{ or }}
\newcommand{\nin}{\not\in}
\newcommand{\inc}{\subseteq}
\newcommand{\set}[1]{\{#1\}}
\newcommand{\setc}[2]{\set{#1 \mid #2}}
\newcommand{\faux}{{\bf F}}
\newcommand{\et}{\wedge}
\newcommand{\ou}{\vee}
\newcommand{\vide}{\emptyset}
\newcommand{\si}{\Leftarrow}
\renewcommand{\iff}{\Leftrightarrow}
\newcommand{\qqs}[2]{\forall\, #1\;\: #2}
\newcommand{\xst}[2]{\exists\, #1\;\: #2}
\newcommand{\nxst}[2]{\not\!\exists\, #1\;\: #2}
\newcommand \seql[3]{\raisebox{3ex}{$\mbox{#1}\;\;$} \; \shortstack{$#2$ \\ \mbox{}\\
                    \mbox{}\hrulefill\mbox{}\\ \mbox{}\\ $#3$}}
\newcommand \seq[2]{\shortstack{$#1$ \\ \mbox{}\\
                    \mbox{}\hrulefill\mbox{}\\ \mbox{}\\ $#2$}}

\newcommand{\forces}{\makebox[5mm]{\,$\|\!-$}}%

\newcommand{\fun}{\:\rightarrow\:} 
\newcommand{\funt}{\rightarrow^{\ast}} 
\newcommand{\lfun}[1]{\stackrel{#1}{\longrightarrow}}
\newcommand{\lfunt}[1]{\stackrel{#1}{\longrightarrow}\,\!^{\ast}}
\newcommand{\length}[1]{ | #1 |}
\newcommand{\union}{\cup}		
\newcommand{\inter}{\cap}		
\newcommand{\Union}{\bigcup}		
\newcommand{\Inter}{\bigcap}		
\newcommand{\join}{\vee}		
\newcommand{\JOIN}{\bigvee}		
\newcommand{\meet}{\wedge}		
\newcommand{\MEET}{\bigwedge}		

\newcommand{\inv}[1]{#1^{-1}}
\newcommand{\card}{\sharp}		
\newcommand{\pfun}{\rightharpoonup}  
\newcommand{\ul}[1]{\underline{#1}}	

\newcommand{\id}{{\it id}}
\newcommand{\catch}{{\it catch}\;}
\newcommand{\throw}{{\it throw}\;}
\newcommand{\Alt}{ \mid\!\!\mid  } 
\newcommand{\Sub}[3]{#1[#2\leftarrow #3]}	
\newcommand{\sub}[2]{[#2 / #1]}
\newcommand{\subn}[2]{[#2 /_{\!\circ} #1]}
\newcommand{\BO}[4]{#1\,,#2\stackrel{#3}{\Longrightarrow} #4}
\newcommand{\bt}[1]{{\it BT}(#1)}
\newcommand{\lelong}[2]{#1\!\mid_{#2}}	
\newcommand{\dl}{[\![} 			
\newcommand{\dr}{]\!]} 			
\newcommand{\newv}{{\it new}\:}
\newcommand{\deref}{{\it deref}\:}
\newcommand{\nat}{{\bf int}}
\newcommand{\varimp}{{\bf var}\:}
\newcommand{\comm}{{\bf comm}}
\newcommand{\world}{{\bf W}}
\newcommand{\comp}{\circ}
\newcommand{\run}{{\it run}}
\newcommand{\done}{{\it done}}
\newcommand{\readv}{{\it read}}
\newcommand{\writev}[1]{{\it write}(#1)}
\newcommand{\OK}{{\it OK}}

\newcommand{\cat}[1]{{\bf #1}}	
\newcommand{\limpl}{\multimap}
\newcommand{\mif}{\mbox{ if }}
\newcommand{\valof}{{\it valof}\:}
\newcommand{\otpt}{{\it output}\:}
\newcommand{\request}{{\it request}\:}
\newcommand{\is}{{\it is}\:}
\newcommand{\pcf}{\mbox{PCF}}
\newcommand{\appl}{\mbox{\raise.28ex\hbox{\tiny $\bullet$}}}
\newcommand{\statecoh}[1]{{\cal C}(#1)}
\newcommand{\play}[2]{#1\ \raisebox{-4pt}{\rule{0.5pt}{4.5mm}} \  #2}
\newcommand{\plays}[2]{#1 \:\mbox{\boldmath $\triangleleft$}\!\ \raisebox{-4pt}{\rule{0.5pt}{4.5mm}} \  #2}
\newcommand{\playc}[2]{#1 \ \raisebox{-4pt}{\rule{0.5pt}{4.5mm}} \ \!\mbox{\boldmath $\triangleright$}\: #2}
\newcommand{\wins}[2]{#1 \mbox{\boldmath $\triangleleft$} #2}
\newcommand{\winc}[2]{#1 \mbox{\boldmath $\triangleright$} #2}
\newcommand{\rst}{\lceil}
\newcommand{\minus}{\backslash}	
\newcommand{\cpt}[1]{{\cal K}(#1)}
\newcommand{\IFF}{\mbox{~~iff~~}}
\newcommand{\dcl}{\downarrow}
\newcommand{\lpar}{\bindnasrepma}

\begin{abstract}
We offer a short tour into the interactive interpretation of sequential programs. We emphasize 
{\em streamlike} computation -- that is, computation of successive bits of information upon request.
The core of the approach surveyed here dates back to the work of Berry and the author on {\em sequential algorithms on concrete data structures} in the late seventies, culminating in the design of the programming language CDS, in which the semantics of programs of any type can be explored interactively.  Around one decade later, two major insights of Cartwright and Felleisen on one hand, and of Lamarche on the other hand gave  new, decisive impulses to the study of sequentiality. Cartwright and Felleisen observed that sequential algorithms give a direct semantics to control operators  like  {\tt call-cc} and proposed to include explicit errors both in the syntax and in the semantics of the language PCF. Lamarche (unpublished) connected sequential algorithms to linear logic and games. The successful program of 
games semantics has spanned over the nineties until now, starting with
 syntax-independent characterizations of the term model of PCF by Abramsky, Jagadeesan, and Malacaria on one hand, and by Hyland and Ong on the other hand.

Only a basic acquaintance with $\lambda$-calculus, domains and linear logic is assumed in sections \ref{prologue} through \ref{SA}.
\end{abstract}

\section{Prologue: playing with B\"ohm trees} \label{prologue}

We first make some preparations. For self-containedness, we briefly recall the relevant notions.
The syntax of the untyped 
$\lbd$-calculus\index{$\lbd$-calculus} ($\lbd$-calculus for short) is 
given by the following three constructions: a {\em variable} $x$ is a $\lbd$-term, if $M$ and $N$ are $\lbd$-terms, then the {\em application} $MN$ is a $\lbd$-term, and if $M$ is a term then the {\em abstraction} $\lbd x.M$ is a term.  Usual abbreviations are $\lbd x_1 x_2.M$ for $\lbd x_1. (\lbd x_2.M)$, and $MN_1N_2$ for $(MN_1)N_2$, and similarly for $n$-ary abstraction and application.
A more macroscopic view is quite useful: it is easy to check that
any $\lbd$-term has exactly one of the following two 
forms:
$$\begin{array}{lll}
(n\geq 1, p\geq 1) && \lbd x_1\cdots x_n.xM_1\cdots M_p\\
(n\geq 0, p\geq 1) &&
\lbd x_1\cdots x_n.(\lbd x.M)M_1\cdots 
M_p
\end{array}$$

\noindent
The first form is called a {\em head normal form} (hnf), while the second exhibits the {\em head redex} 
$ (\lbd x.M)M_1$. The following easy property justifies the name of head normal form:
any reduction sequence starting from a hnf $\lbd x_1\cdots 
x_n.xM_1\cdots M_p$  consists of an interleaving of independent 
reductions of 
$M_1,\ldots,M_p$. More precisely, we have:
 $$\begin{array}{l}
(\lbd x_1\cdots x_n.xM_1\cdots M_p\funt P)\Rightarrow \\\xst N_1,\ldots 
N_p{\left\{\begin{array}{l}
P=\lbd x_1\cdots x_n.xN_1\cdots N_p\mand\\
\qqs{i\leq p}{M_i}\funt N_i\; .
\end{array}\right.}
\end{array}$$

\noindent
Here, reduction means the replacement in any term of a sub-expression  of the form $(\lbd x.M)N$, called a $\beta$-redex,
by $\Sub{M}{x}{N}$. 
A normal form is a term that contains no $\beta$-redex, or equivalently that contains no head redex. Hence the syntax of normal forms is given by the following two constructions: a variable $x$ is a normal form, and if $M_1,\ldots,M_p$ are normal forms, then $ \lbd x_1\cdots x_n.xM_1\cdots M_p$ is a normal form.

\medskip
Now, we are ready to play.  Consider the following two normal forms:
$$M= zM_1M_2(\lbd z_1z_2.z_1M_3M_4)\quad\quad N= \lbd x_1x_2 x_3. x_3( \lbd y_1y_2.y_1N_1) N_2$$
The term $\Sub{M}{z}{N}=NM_1M_2(\lbd z_1z_2.z_1M_3M_4)$ is not a normal form anymore, and can be $\beta$-reduced as follows:
$$\begin{array}{lll}
NM_1M_2(\lbd z_1z_2.z_1M_3M_4) & \rightarrow & (\lbd z_1z_2.z_1M_3M_4)( \lbd y_1y_2.y_1N'_1) N'_2\\
& \rightarrow & ( \lbd y_1y_2.y_1N'_1)M'_3M'_4 \\
& \rightarrow & M'_3N''_1
\end{array}$$
where $N'_1$, etc... are suitable substitution instances of $N_1$ etc...
But there is a more geometric way of describing  the {\em interaction} of $M$ and $N$. First, we represent $M$ and $N$ explicitly as trees (read from left to right), as follows:

$$\begin{array}{cccc}
z\begin{branch}M_1\\
M_2\\
\lbd z_1z_2.\;z_1\begin{branch} M_3\\
M_4\end{branch}\end{branch}
&&
\lbd x_1x_2x_3.\;x_3
\begin{branch}\lbd y_1y_2.\;y_1\begin{branch}N_1\end{branch}\\N_2
\end{branch}\end{array}$$

\noindent
Then we represent computation as the progression of two tokens in the two trees. Initially, the tokens are at the root (we use underlining to indicate the location of the tokens):

$$\begin{array}{cccc}
\ul{z}\begin{branch}M_1\\
M_2\\
\lbd z_1z_2.\;z_1\begin{branch} M_3\\
M_4\end{branch}\end{branch}
&&
\ul{\lbd x_1x_2 x_3.}\;x_3
\begin{branch}\lbd y_1y_2.\;y_1\begin{branch}N_1\end{branch}\\N_2
\end{branch}\end{array}$$

\noindent
We then notice that the token in $M$ has a choice of where to move right, while the one on the right has no choice. So let us take the latter as pilot:

$$\begin{array}{cccc}
\ul{z}\begin{branch}M_1\\
M_2\\
\lbd z_1z_2.\;z_1\begin{branch} M_3\\
M_4\end{branch}\end{branch}
&&
\lbd x_1x_2 x_3.\;\ul{x_3}
\begin{branch}\lbd y_1y_2.\;y_1\begin{branch}N_1\end{branch}\\N_2
\end{branch}\end{array}$$

\noindent
Here, $\ul{x_3}$ reads as ``take the third branch'' (in $M$):

$$\begin{array}{cccc}
z\begin{branch}M_1\\
M_2\\
\ul{\lbd z_1z_2.}\;z_1\begin{branch} M_3\\
M_4\end{branch}\end{branch}
&&
\lbd x_1x_2 x_3.\;\ul{x_3}
\begin{branch}\lbd y_1y_2.\;y_1\begin{branch}N_1\end{branch}\\N_2
\end{branch}\end{array}$$

\noindent
Notice that we are now in a situation where the token in $M$ knows how to move next, while the one in $N$ has a choice. We take $M$ as pilot:

$$\begin{array}{cccc}
z\begin{branch}M_1\\
M_2\\
\lbd z_1z_2.\;\ul{z_1}\begin{branch} M_3\\
M_4\end{branch}\end{branch}
&&
\lbd x_1x_2 x_3.\;\ul{x_3}
\begin{branch}\lbd y_1y_2.\;y_1\begin{branch}N_1\end{branch}\\N_2
\end{branch}\end{array}$$

\noindent 
We read $z_1$ as ``take the first branch'' (in $N$):

$$\begin{array}{cccc}
z\begin{branch}M_1\\
M_2\\
\lbd z_1z_2.\;\ul{z_1}\begin{branch} M_3\\
M_4\end{branch}\end{branch}
&&
\lbd x_1x_2 x_3.\;x_3
\begin{branch}\ul{\lbd y_1y_2.}\;y_1\begin{branch}N_1\end{branch}\\N_2
\end{branch}\end{array}$$

\noindent
The following steps are:

$$\begin{array}{cccc}
z\begin{branch}M_1\\
M_2\\
\lbd z_1z_2.\;\ul{z_1}\begin{branch} M_3\\
M_4\end{branch}\end{branch}
&&
\lbd x_1x_2 x_3.\;x_3
\begin{branch}\lbd y_1y_2.\;\ul{y_1}\begin{branch}N_1\end{branch}\\N_2
\end{branch}\end{array}$$

$$\begin{array}{cccc}
z\begin{branch}M_1\\
M_2\\
\lbd z_1z_2.\;z_1\begin{branch} \ul{M_3}\\
M_4\end{branch}\end{branch}
&&
\lbd x_1x_2 x_3.\;x_3
\begin{branch}\lbd y_1y_2.\;\ul{y_1}\begin{branch}N_1\end{branch}\\N_2
\end{branch}\end{array}$$

\noindent
We leave it to the reader to check that these steps follow closely the sequence of $\beta$-reductions given above. The graphical presentation enhances alternation. The tokens' moves alternate between $M$ and $N$. There are two sorts of moves: variables (like $z$), and (bunches of) abstractions (like $\lbd z_1z_2.$). We call these moves Player's moves and Opponent's moves, respectively. We can view an Opponent's move as the question ``what is the head variable of the term rooted here?'', and a Player's move as the answer to this question. 
So we see the computation as a progression of alternating moves describing a path in $M$ (and in $N$):

$$\begin{array}{cccc}
\ul{z}\begin{branch}M_1\\
M_2\\
\ul{\lbd z_1z_2.}\;\ul{z_1}\begin{branch} \ul{M_3}\\
M_4\end{branch}\end{branch}
&&
\ul{\lbd x_1x_2 x_3.}\;\ul{x_3}
\begin{branch}\ul{\lbd y_1y_2.}\;\ul{y_1}\begin{branch}N_1\end{branch}\\N_2
\end{branch}\end{array}$$

\noindent
Our example is actually too simple. The general mechanism needs an explicit manipulation of pointers, when (unlike in the example) a variable is not bound by the immediate bunch of $\lambda$'s above. We refer the interested reader to \cite{ABT,CH97}, where this machinery is described for a larger class of trees with pointers -- called abstract B\"ohm trees --, of which B\"ohm trees are an example. Our main point here was to highlight interaction:  $M$ and $N$ are pilot in turn and tell the other which branch to choose. 

\medskip
Suppose now that $M_3=\lbd u.tM_5M_6$, where $t$ is a free variable of $\Sub{M}{z}{N}$. Then, looking back at the sequence of $\beta$-reductions, we reach a head normal form:
$$\begin{array}{lll}
M'_3N''_1= (\lbd u.tM'_5M'_6)N''_1 & \rightarrow & 
tM''_5M''_6
\end{array}$$
And, in geometrical form:

$$\begin{array}{cccc}
z\begin{branch}M_1\\
M_2\\
\lbd z_1z_2.\;z_1\begin{branch}\lbd u.\;\ul{t}\begin{branch}M_5\\M_6\end{branch}\\
M_4\end{branch}\end{branch}
&&
\lbd x_1x_2 x_3.\;x_3
\begin{branch}\lbd y_1y_2.\;\ul{y_1}\begin{branch}N_1\end{branch}\\N_2
\end{branch}\end{array}$$

\noindent
Note here that $N$ cannot help to choose the next move in $M$. The machinery stops here. After all, most functional programming languages stop evaluation on (weak) head normal forms.
But what about getting the full normal form, i.e., computing $M''_5$ and $M''_6$? The interactive answer to this question is: by exploration of branches, on demand, or in a streamlike manner.
The machine displays $t$ as the head variable of the normal form of $\Sub{M}{z}{N}$. Now, you, the Opponent, can choose which of the branches below $t$ to explore, and then the machine will restart until it reaches a head normal form. For example, if you choose the first branch, then you will eventually reach the head variable of $M''_5$. This is called streamlike, because that sort of mechanism has been first analysed for infinite lists built progressively. A $\lbd$-term too has a ``potentially infinite normal form'': it's B\"ohm tree.

This prologue served the purpose of introducing some keywords, such as interactivity, playful interpretation, streamlike computation. We now start from the beginning.

\section{Introduction} \label{intro}

Scott's and Plotkin's denotational semantics takes its roots in recursion theory. It is worth recalling here the statement of Rice's theorem. This theorem asserts a property of recursively enumerable (r.e.) sets of partial recursive (p.r.) functions, defined through a fixed enumeration $(\phi_n)$ of the p.r. functions (i.e. $\phi$ is a surjection from $\omega$ -- the set of natural numbers -- to $\omega\rightharpoonup\omega$, using $\rightharpoonup$ for sets of partial functions). Let $PR\subseteq \omega\rightharpoonup\omega$ denote the set of p.r. functions. A subset $A\inc PR$ is called r.e. if $\setc{n}{\phi_n\in A}$ is r.e. in the usual sense. The theorem asserts that if $A$ is r.e. and if $f\in A$, then there exists a finite approximation $g$ of $f$ such that $g\in A$.
That $g$ is an approximation of $f$ means that $f$ is an extension of $g$, i.e., the domain on which the partial function $f$ is defined, or domain of definition of $f$, contains that of $g$ and $f$ and $g$ coincide where they are both defined. A simpler way of saying this is that the graph of $g$ is contained in the graph of $f$.
Moreover, the domain of definition of $g$ is finite.  Rice's theorem is about an intrinsic continuity property in the realm of p.r. functions. It highlights the (complete) partial order structure of 
$\omega\rightharpoonup\omega$, and in particular the presence of a bottom element $\bot$ in this partial order: the everywhere undefined function.

Certainly, one of the key departure points taken by Scott was to take $\bot$ seriously. Once this element is part of the picture, one takes a new look at some basic functions. Take the booleans, for example. In Scott's semantics, this is not the set $\set{{\it tt},{\it ff}}$, but the set  $\set{\bot,{\it tt},{\it ff}}$ ordered as follows: $x\leq y$ if and only if $x=y$ or $x=\bot$ (this is called {\em flat} ordering). Take now the good old disjunction function ${\it or}:Bool\times Bool\rightarrow Bool$. It gives rise to four different functions over the flat domain version of $Bool$ (the specifications below can be completed to full definitions by monotonicity):

$$\begin{array}{l}
por(\bot,{\it tt}) ={\it tt}\\ 
por({\it tt},\bot)={\it tt}\\
por(\bot,{\it ff})=\bot\\
por({\it ff},\bot)=\bot\\
por({\it ff},{\it ff})={\it ff}
\end{array}$$

$$\begin{array}{l}
lor(\bot,y) =\bot\\ 
lor({\it tt},\bot)={\it tt}\\
lor({\it ff},y)=y\\
\end{array} \quad\quad\quad\quad
\begin{array}{l}
ror(\bot,{\it tt}) ={\it tt}\\ 
ror(x,\bot)=\bot\\
ror(x,{\it ff})=x
\end{array}$$

$$\begin{array}{l}
sor(\bot,{\it tt}) =\bot\\ 
sor({\it tt},\bot)=\bot\\
sor(\bot,{\it ff})=\bot\\
sor({\it ff},\bot)=\bot\\
sor({\it ff},{\it tt})={\it tt}\\
sor({\it tt},{\it ff})={\it tt}\\
sor({\it tt},{\it tt})={\it tt}\\
sor({\it ff},{\it ff})={\it ff}
\end{array}$$

\medskip
It should be clear that $lor$ and $ror$ are computed by  programs of the following shape, respectively:

$$\lbd xy.\:{\tt if}\:x={\it tt}\left\{\begin{array}{l}
{\tt then}\:{\it tt}\\
{\tt else}\:{\tt  if}\:y=\cdots
\end{array}\right.\quad\quad\quad\quad
\lbd xy.\:{\tt if}\:y={\it tt}\left\{\begin{array}{l}
{\tt then}\:{\it tt}\\
{\tt else}\:{\tt if}\:x=\cdots
\end{array}\right.$$

\medskip 
On the other hand, it should be intuitively clear that no sequential program of the same sort can compute $por$, because a sequential program will 
\begin{itemize}
\item either start by examining one of the arguments, say $x$, in which case it can't output  anything before a value for $x$ is given, thus missing the specification $por(\bot,{\it tt})={\it tt}$,
\item or output some value rightaway, say ${\it tt}$  ($\lbd xy.{\it tt}$), thus mising the specification 
$por(\bot,\bot)=\bot$.
\end{itemize}

For a formal proof that $por$ is not sequentially definable, we refer to \cite{Plo77} (syntactic proof), to \cite{Gunter92}[section 6.1] (model-theoretic proof), and to \cite{AmaCu}[section 4.5] (via logical relations). As for $sor$, the story is yet different, there are {\em two} natural sequential programs for it:

$$\lbd xy.\:{\tt if}\:x={\it tt}\left\{\begin{array}{l}
{\tt then}\:{\tt  if }\:y=\cdots\\
{\tt else}\:{\tt  if}\:y=\cdots
\end{array}\right.\quad\quad\quad\quad
\lbd xy.\:{\tt if}\:y={\it tt}\left\{\begin{array}{l}
{\tt then}\:{ \tt if}\:x=\cdots\\
{\tt else}\:{\tt  if}\:x=\cdots
\end{array}\right.$$

The starting point of the model of sequential algorithms (next section) was to interpret these two programs as different objects ${\it lsor}$ and ${\it rsor}$.
Notice finally that there are many more sequential programs computing $lor$, $ror$, or $sor$.
Another program for $lor$ might e.g.  look like

$$\lbd xy.\:{\tt if}\:x={\it tt}\left\{\begin{array}{l}
{\tt then}\:{\it tt}\\
{\tt else}\:{\tt  if}\:x={\it tt}\left\{\begin{array}{l}
{\tt then}\:{\it tt}\\
{\tt else}\:{\tt  if}\: y=\cdots
\end{array}\right.
\end{array}\right.$$

Such a ``stuttering'' program is perfectly correct syntactically. Whether this program is interpreted in the model by an object different from the above program for $lor$ is the departure point between the model of sequential algorithm on one hand and the more recent games semantics on the other hand. We shall come back to this point in the next section.

Before we close the section, let us give some rationale for the names used in this section. As the reader might have guessed, the prefixes $p,l,r,s,ls,rs$ stand for ``parallel", ``left", ``right", ``left strict", and ``right strict", respectively.

\section{Symmetric algorithms, sequential algorithms}  \label{SA} 

We introduce enough formal definitions to give a self-contained introduction to sequential algorithms, presented in the light of a games interpretation \cite{Symseq} (following work of Lamarche \cite{Lam92}).  The proofs are omitted, but can be found in \cite{AmaCu}[section 14.3], except for what regards the coincidence between the two definitions of composition, for which the proof from \cite{Cur86}[section 3.6] can easily be adapted.

\begin{definition} \label{sds-Seq}
A {\em sequential data structure}\index{sequential data structure 
(sds)}  $\cat{S}=(C,V,P)$ is given by two sets
$C$ and $V$ of {\em cells} and {\em values}, which are assumed disjoint, and by a 
collection $P$ of non-empty words 
$p$ of the form:
$$c_1v_1\cdots c_nv_n\mbox{ or }c_1v_1\cdots c_{n-1}v_{n-1}c_n,$$
where 
$c_i\in C$ and $v_i\in V$ for all $i$. Thus any $p\in P$ is 
alternating and starts with a 
cell. Moreover, it is assumed that $P$ is closed under non-empty 
prefixes.
We call the 
elements of $P$ {\em positions} of $\cat{S}$.
	We call {\em move} any element of $M=C\union V$. 
We use $m$ to denote a move. A position ending with 
a value is called a {\em response}\index{response}, and a position 
ending with a cell is called a {\em query}\index{query}. We use $p$ 
(or $s$, or 
$t$), $q$, and $r$, to range over positions, queries, and responses, 
respectively. We denote by $Q$ and 
$R$ the sets of queries and responses, respectively.
\end{definition}

Let us pause here for some comments and perspective. An important step in the semantic account of sequential computing was taken by Berry, who developed the stable model in which the function $por$ is excluded. Winskel described this model more concretely in terms of {\em event structures}, and Girard proposed a simpler form called {\em coherence spaces}, that led him to the discovery of linear logic \cite{Gir87}  (see also \cite{AmaCu}[chapters 12 and 13]). In event structures or coherence spaces, data are constructed out of elementary pieces, called {\em events}, or tokens. For example, the pair of booleans
$({\it tt},{\it ff})$ is obtained as the set of two elementary pieces: $({\it tt},\bot)$ and $(\bot,{\it ff})$.
More precisely and technically, the structure $Bool\times Bool$ as a coherence space has four events: ${\it tt}.1$,
${\it ff}.1$, ${\it tt}.2$, and ${\it ff}.2$. Then $({\it tt},{\it ff})$ is the set $\set{{\it tt}.1,{\it ff}.2}$.

In a sequential data structure (or in a concrete data structure, not defined here) events are further cut in two ``halves'': a cell and a value, or an opponent's move and a player's move.
The structure $Bool\times Bool$ as an sds has two cells $?.1$ and $?.2$ and has four values
${\it tt}.1$,
${\it ff}.1$, ${\it tt}.2$, and ${\it ff}.2$.  An event, say ${\it tt}.1$, is now decomposed as a position
$(?.1)\:({\it tt}.1)$. The best way to understand this is to think of a streamlike computation.
Your pair of booleans is the output of some program, which will only work on demand. The cell
$?.1$ reads as ``I -- another program, or an observer -- want to know the
 left coordinate of the result of the program'', and ${\it tt}.1$ is the answer to this query.

An important remark, which will be further exploited in section \ref{control}, is that this decomposition of events gives additional space: there is no counterpart in the
world of coherence spaces or in any other usual category of domains of a structure with one cell and no value.
\begin{definition}
A {\em strategy}\index{strategy} of \cat{S} is a subset $x$ of $R$ 
that is closed under response prefixes and binary non-empty greatest lower bounds (glb's):
\begin{eqnarray*}
r_1,r_2\in x, r_1\meet r_2\neq\epsilon & \Rightarrow & r_1\meet r_2\in x
\end{eqnarray*}
where $\epsilon$ denotes the empty word.
A {\em counter-strategy}\index{counter-strategy} is a non-empty 
subset of $Q$ that is closed under query prefixes and under binary 
glb's. We use $x,y,\ldots$ and $\alpha,\beta,\ldots$ to range over 
strategies and counter-strategies, 
respectively.

If $x$ is a strategy and if $r\in x, q=rc\mbox{ for some }c$ and
if there is no $v$ such that $qv\in x$, we write 
$q\in A(x)$
(and say that $q$ is accessible from $x$).
Likewise we define $r\in A(\alpha)$ for a response $r$ 
and a counter-strategy 
$\alpha$.

	Both sets of strategies and of counter-strategies are ordered by 
inclusion. They are denoted by 
$D(\cat{S})$\index{$D(\cat{S})$} and 
$D^{\bot}(\cat{S})$\index{$D^{\bot}(\cat{S})$}, respectively. We write
$\cpt{D(\cat{S})}$ and
$\cpt{D^{\bot}(\cat{S})}$ for the sets of {\em finite} strategies and counter-strategies, respectively.
Notice 
that $D(\cat{S})$ has always a minimum 
element (the empty strategy, written $\emptyset$ or $\bot$), while 
$D^{\bot}(\cat{S})$ has no 
minimum element in general. \end{definition}

A more geometric reading of the definitions of sds, strategy and 
counter-strategy is the following.
An sds is a labelled forest, where the ancestor relation 
alternates cells and values, and where the 
roots are labelled by cells. A strategy is a sub-forest which is allowed to branch only at 
values.
A counter-strategy $\alpha$ is a non-empty subtree which is 
allowed to branch only at cells.

\medskip 
Let us see what collections of positions form and do not form a strategy 
in $Bool\times Bool$.  The set $\set{(?.1)\:({\it tt}.1)\:,\:(?.2)\:({\it ff}.2})$ (representing $({\it tt},{\it ff})$)
is a strategy, while $\set{(?.1)\:({\it tt}.1)\:,\:(?.1)\:({\it ff}.1)}$ is not a strategy. A way to understand this is to say that the cell $?.1$ can hold only one value, which is the answer to the question. A strategy consists in having ready determinate answers for the movements of the opponent.
If strategies are data, what are counter-strategies? They can be considered as exploration trees, see below. 

The pairs cell--value, query--response, and strategy--counter-strategy give to sds's a flavour of 
symmetry. These pairs are related to other important dualities in 
programming: input--output, 
constructor--destructor (see \cite{sym-inter}).
It is thus tempting to consider the counter-strategies of an sds 
\cat{S} as the strategies of a dual 
structure $\cat{S}^{\bot}$ whose cells are the values of \cat{S} 
and whose values are 
the cells of \cat{S}. However, the 
structure obtained in this way is not an sds anymore, since positions 
now start with a value.  This situation, first analysed by Lamarche  \cite{Lamarche92a}, is now well-understood since the thesis work of Laurent \cite{Laurent2002}. We come back to this below.

\medskip
The following definition resembles quite closely to the dynamics described in section \ref{prologue}.
\begin{definition}[play] \label{play-Seq}
Let $\cat{S}$ be an sds, $x$ be a strategy and $\alpha$ be a 
counter-strategy of \cat{S}, one of which 
is finite. We define $\play{x}{\alpha}$\index{$\play{x}{\alpha}$}, 
called a {\em play}\index{play}, as the set of positions $p$ which 
are such that all the response 
prefixes of $p$ are in $x$ and all the query prefixes of $p$ are in 
$\alpha$.
\end{definition}
\begin{proposition} Given $x$ and $\alpha$ as in definition 
\ref{play-Seq},
the play $\play{x}{\alpha}$ 
is non-empty and totally ordered, and can be confused with its maximum element, 
which is uniquely characterized as 
follows:

\medskip
$\begin{array}{ll}
\play{x}{\alpha}\mbox{ is the unique element of }x\inter A(\alpha) 
&\mif \play{x}{\alpha}\mbox{ is 
a response}\\
\play{x}{\alpha}\mbox{ is the unique element of }\alpha\inter A(x) & 
\mif \play{x}{\alpha}\mbox{ is a 
query}\; .
\end{array}$
\end{proposition}
\begin{definition}[winning]
Let $x$ and $\alpha$ be as in definition \ref{play-Seq}.
If $\play{x}{\alpha}$ is a response, we say that $x$ wins against 
$\alpha$, and we denote this 
predicate by $\wins{x}{\alpha}$\index{$\wins{x}{\alpha}$}. If 
$\play{x}{\alpha}$ is a query, we say that $\alpha$ wins against 
$x$, and we write $\winc{x}{\alpha}$\index{$\winc{x}{\alpha}$}, thus 
$\winc{}{}$ is the negation of $\wins{}{}$. To stress 
who is the winner\index{$\playc{x}{\alpha}$}, we 
write\index{$\plays{x}{\alpha}$}:
$$\play{x}{\alpha}=\left\{\begin{array}{ll}
\plays{x}{\alpha} &\mbox{when }x\mbox{ wins}\\
\playc{x}{\alpha} &\mbox{when }\alpha\mbox{ wins}\; .
\end{array}\right.$$
\end{definition}

	The position $\play{x}{\alpha}$ formalizes the interplay between the 
player with strategy $x$ and the 
opponent with strategy $\alpha$. If $\play{x}{\alpha}$ is a response, 
then the player wins since he 
made the last move, and if $\play{x}{\alpha}$ is a query, then the 
opponent wins. Here is a game theoretical reading of 
$\play{x}{\alpha}$. At the beginning 
the opponent makes a move $c$: his strategy 
determines that move uniquely. Then either the player is unable to 
move ($x$ contains no position of the 
form $cv$), or his strategy determines a unique move. The play goes 
on until one of $x$ or $\alpha$ 
does not have the provision to answer its opponent's move (cf. section \ref{prologue}).

\medskip
We next define the morphisms between sds's. There are two definitions, a concrete one and a more abstract one. The concrete one is needed since we want the morphisms to form in turn an sds in order to get a cartesian closed category (actually a monoidal closed one, to start with).  
Accordingly, there will be two definitions of the composition of morphisms. Their equivalence is just what {\em full abstraction} --  that is, the coincidence of operational and denotational semantics -- boils down to, once we have tailored the model to the syntax (programs as morphisms) and tailored the syntax to the semantics (like in the language CDS \cite{BerryCurien85}).
We start with the concrete way.

\begin{definition}
Given sets $A,B\inc A$, for any word $w\in A^{\ast}$, we define 
$w\rst_B$\index{$w\rst_B$} as follows:
$$\epsilon\rst_B=\epsilon \quad\quad
wm\rst_B = \left\{\begin{array}{ll}
w\rst_B & \mif m\in A\minus B\\
(w\rst_B)m & \mif m\in B\; .
\end{array}\right.$$
\end{definition}
\begin{definition} 
\label{affine-exponent-Seq}
Given two sds's $\cat{S}=(C,V,P)$ and $\cat{S}'=(C',V',P')$, we define 
$\cat{S}\limpl \cat{S}'=(C'',V'',P'')$ as follows. The sets
$C''$ and $V''$ are disjoint unions:
$$\begin{array}{lll}
C'' & = & \setc{\request c'}{c'\in C'}\union\setc{\is v}{v\in V}\\
V'' & = & \setc{\otpt v'}{v'\in V'}\union\setc{\valof c}{c\in C}\; .
\end{array}$$
$P''$ consists of the alternating positions $s$ starting with a 
$\request c'$, and which are such that:
$$\begin{array}{l}
s\rst_{\cat{S}'}\in P', (s\rst_\cat{S}=\epsilon\mor s\rst_\cat{S}\in 
P), \mand\\
s\mbox{ has no prefix of the form }s(\valof c)(\request c').
\end{array}$$
We often omit the tags ${\it request},{\it valof},{\it is},{\it output}$, 
as we have just 
done
in the notation $s\rst_\cat{S}=s\rst_{C\union V}$ (and similarly for
$s\rst_{\cat{S}'}$).

We call {\em affine sequential algorithms}\index{affine algorithm} 
(or affine algorithms) from \cat{S} to $\cat{S}'$ the strategies of 
$\cat{S}\limpl\cat{S}'$. 
\end{definition}

The constraint `no $scc'$' can be formulated more informally as 
follows. Thinking of $\valof c$ as a 
call to a subroutine, the principal routine cannot proceed further 
until it receives a result $v$ from the 
subroutine.

\medskip
The identity affine algorithm $\id\in 
D(\cat{S}\limpl\cat{S}')$ is defined as 
follows:
$$\id=\setc{{\it copycat}(r)}{r\mbox{ is a response of }\cat{S}},$$
where ${\it copycat}$ is defined as follows:
$$\begin{array}{lll}
{\it copycat}(\epsilon)& = & \epsilon \\
{\it copycat}(rc) & = & {\it copycat}(r)(\request c)(\valof c) \\
{\it copycat}(qv) & = & {\it copycat}(q)(\is v)(\otpt v)\; .
\end{array}$$
The word ${\it copycat}$ used in the description of the identity 
algorithm has been proposed in \cite{Abr-Jag-92}, and corresponds to a 
game theoretical understanding: the player always repeats the last move 
of the opponent. In some influential talks, Lafont had taken images from chess (Karpov -- Kasparov) to explain the same thing.
	
\begin{example} \label{affine-algo-ex-Seq}
$(1)$ The following  affine algorithm 
computes the boolean negation 
function:
$$\begin{array}{l}
\{(\request ?)(\valof ?),\\
(\request ?)(\valof ?)(\is {\it tt})(\otpt {\it ff}),\\
(\request ?)(\valof ?)(\is {\it ff})(\otpt {\it tt})\}\; .
\end{array}$$
$(2)$ On the other hand, the left disjunction function  cannot be 
computed
by an affine algorithm. Indeed, transcribing the program for $lor$ as a strategy leads to:
$$\begin{array}{l}\{(\request ?)(\valof ?.1),\\ 
(\request ?)(\valof ?.1)(\is {\it tt})(\otpt {\it tt}),\\
(\request ?)(\valof ?.1)(\is {\it ff})(\valof ?.2),\\
(\request ?)(\valof ?.1)(\is {\it ff})(\valof ?.2)(\is {\it tt})(\otpt 
{\it tt}),\\
(\request ?)(\valof ?.1)(\is {\it ff})(\valof ?.2)(\is {\it ff})(\otpt 
{\it ff})\}\; ,
\end{array}$$ 
which is not a subset of the set of positions of 
$Bool^2\limpl Bool$, because the projections on 
$Bool^2$ of the last two sequences of moves are not positions of 
$Bool^2$.  But the program does transcribe into a (non-affine) sequential algorithm, as we shall see.

\medskip\noindent
(3) Every constant function gives rise to an affine algorithm, whose
responses have the form
$(\request c'_1)(\otpt v'_1)\ldots (\request c'_n)(\otpt v'_n).$.

\end{example}

The second and third example above thus justify the terminology affine (in the affine framework, in contrast to the linear one, weakening is allowed).
The second example suggests the difference between 
affine and general sequential 
algorithms. Both kinds of algorithms ask successive queries to their 
input, and continue to proceed only after they 
get responses to these queries. An affine algorithm is moreover required 
to ask these queries 
monotonically: each new query must be an extension of the previous 
one. The `unit' of resource 
consumption 
is thus a sequence of queries/responses that can be 
arbitrarily large, as long as it builds a 
position of the input sds. The disjunction algorithms are not affine, 
because they may have to ask 
successively the queries $?.1$ and $?.2$, which are not related by 
the prefix ordering. 

A generic affine algorithm, as represented in figure 
\ref{generic-aff-Seq},  can be viewed as a 
`combination' of the following (generic) output strategy and input  
counter-strategy (or exploration tree):
$$\begin{array}{cccc}
\mbox{input counter-strategy} & & & \mbox{output strategy}\\
c\left\{\begin{array}{l}
v_1\;\cdots\\
\vdots\\
v_i\; d\left\{\begin{array}{l}
\vdots\\
w\\
\vdots\\
\end{array}\right.\\
\vdots\\
v_n\;\cdots
\end{array}\right. && &
c'\; v'\left\{\begin{array}{l}
c'_1\;\cdots\\
\vdots\\
c'_m\;\cdots
\end{array}\right.
\end{array}$$

\begin{figure}
{\small
$$\request c'\; \valof c\left\{\begin{array}{l}
\is v_1\;\cdots\\
\vdots\\
\is v_i\;\valof d\left\{\begin{array}{l}
\vdots\\
\is w\;\otpt v'\left\{\begin{array}{l}
\request c'_1\;\cdots\\
\vdots\\
\request c'_m\;\cdots
\end{array}\right.\\
\vdots
\end{array}\right.\\
\vdots\\
\is v_n\;\cdots
\end{array}\right.$$}

\caption{A generic affine algorithm}
\label{generic-aff-Seq}
\end{figure}

We now give a definition of composition of affine algorithms by means of a simple {\em abstract machine}. Sequential algorithms are syntactic objects, and were indeed turned into a programming language called CDS \cite{BerryCurien85}. What we present here is a simplified version of the operational semantics presented in \cite{Cur86}[section 3.5] in the special case of affine algorithms. Given $\phi\in D(\cat{S}\limpl\cat{S}')$ and $\phi'\in D(\cat{S}'\limpl\cat{S}'')$, the goal is to compute on demand the positions that belong to their composition $\phi''$ in the sds $\cat{S}\limpl\cat{S}''$. The abstract machine proceeds by rewriting triplets $(s,s',s'')$ where $s,s',s''$ are positions of $\cat{S}\limpl\cat{S}'$, $\cat{S}'\limpl\cat{S}''$, and $\cat{S}\limpl\cat{S}''$, respectively. The rules are given in Figure \ref{abstract-machine} (where $P''$ designates the set of positions of 
$\cat{S}\limpl\cat{S}''$, etc...):

The first two rules are left to the (streamlike)  initiative of the observer. Each time one of these rules is activated, it
 launches the machine proper, that consists of the four other (deterministic) rules.
The generic behaviour of the machine is as follows. The initial triplet is $(\epsilon,\epsilon,\epsilon)$.  The observer wants to know the content of $c''$, or more precisely wants to know what the function does in order to compute the contents of $c''$ in the output. Thus, he chooses to perform the following rewriting:
$$\begin{array}{lll}
(\epsilon,\epsilon,\epsilon) & \longrightarrow & (\epsilon,\epsilon,c'')
\end{array}$$
The request is transmitted to $\phi'$:
$$\begin{array}{lll}
(\epsilon,\epsilon,c'') & \longrightarrow & (\epsilon,c'',c'')
\end{array}$$
There are two cases here. Either $\phi'$ does not consult its input and produces immediately a value for $c''$, in which case, this value is transmitted as the final result of the oberver's query:
$$\begin{array}{lllll}
(\epsilon,c'',c'') & \longrightarrow & (\epsilon,c''v'',c''v'') && (c''v''\in\phi')
\end{array}$$
Or $\phi'$ needs to consult its input (like the various sequential or functions), and then an
{\em interaction loop} (in the terminology of Abramsky and Jagadeesan \cite{AJ92}) is initiated:
$$\begin{array}{lllll}
(\epsilon,c'',c'') & \longrightarrow & (c'_1,c''c'_1,c'') && (c''c'_1\in\phi')\\
& \longrightarrow & (c'_1v'_1,c''c'_1v'_1,c'') && (c'_1v'_1\in\phi)\\
& \longrightarrow & (c'_1v'_1c'_2,c''c'_1v'_1c'_2,c'') && (c''c'_1v'_1c'_2\in\phi')\\
\vdots
\end{array}$$

\begin{figure}
$$\begin{array}{lllll}
(r,r',r'') & \longrightarrow & (r,r'c'',r''c'') && (r''c''\in P'')\\
(r,r',r'') & \longrightarrow & (rv,r'c'',r''v) && (r''v\in P'')\\\\
(r,q',q'') & \longrightarrow & (r,q'v'',q''v'') && (q'v''\in\phi')\\
(r,q',q'') & \longrightarrow & (rc',q'c',q'') && (q'c'\in\phi')\\
(q,r',q'') & \longrightarrow & (qv',r'v',q'') && (qv'\in\phi)\\
(q,r',q'') & \longrightarrow & (qc,r',q''c) && (qc\in\phi)
\end{array}$$
\caption{Composition abstract machine for affine algorithms}
\label{abstract-machine}
\end{figure}

This dialogue between $\phi$ and $\phi'$ may terminate in two ways. Either at some stage
 $\phi'$ has received enough information from $\phi$ to produce a value $v''$ for $c''$, i.e. $c'_1v'_1\ldots c'_nv'_nv''\in\phi'$:
$$\begin{array}{lll}
 (c'_1v'_1\ldots c'_nv'_n,c''c'_1v'_1c'_2\ldots c'_nv'_n,c'')
& \longrightarrow & (c'_1v'_1\ldots c'_nv'_n,c''c'_1v'_1c'_2\ldots c'_nv'_nv'',c''v'') 
\end{array}$$
or
$\phi$ itself says it needs to consult its input, i.e.,  $c'_1v'_1\ldots c'_nc\in\phi$: this information is passed as a final (with respect to the query $c''$) result to the observer, who then knows that $\phi''$ needs to know the content of $c$. 
$$\begin{array}{lll}
 (c'_1v'_1\ldots c'_n,c''c'_1v'_1c'_2\ldots c'_n,c'')
& \longrightarrow & (c'_1v'_1\ldots c'_nc,c''c'_1v'_1c'_2\ldots c'_n,c''c) 
\end{array}$$
It is then the observer's freedom to explore further the semantics of $\phi''$ by issuing a new query (provided it is in $P''$) :
$$\begin{array}{lllll}
 (c'_1v'_1\ldots c'_nv'_n,c''c'_1v'_1c'_2\ldots c'_nv'_nv'',c''v'') 
& \longrightarrow &  (c'_1v'_1\ldots c'_nv'_n,c''c'_1v'_1c'_2\ldots c'_nv'_nv'',c''v''c''_1)  && 
\end{array}$$
or
$$\begin{array}{lllll}
(c'_1v'_1\ldots c'_nc,c''c'_1v'_1c'_2\ldots c'_n,c''c)
& \longrightarrow & (c'_1v'_1\ldots c'_nc,c''c'_1v'_1c'_2\ldots c'_n,c''cv) && 
\end{array}$$
The query $c''cv$ reads as: ``knowing that $\phi''$ needs $c$, how does it behave next when I feed $v$ to $c$''.  After this, the computation starts again using the four deterministic rules along the same general pattern. Notice how $\phi$ and $\phi'$ take in turn  the leadership  in the interaction loop (cf. section \ref{prologue}.

\medskip
We now turn to the abstract definition of our morphisms.

\begin{definition}
A (continuous) function 
$f:D(\cat{S})\fun D(\cat{S}')$ is called
{\em stable} if for any $x\in 
D(\cat{S}),\alpha'\in\cpt{D^{\bot}$ such that $\cat{S}')},\wins{f(x)}{\alpha'}$ there exists a minimum (finite) $y\leq x$ such that 
$\wins{f(y)}{\alpha'}$ ($m(g,\alpha',x)$, denoted by $m(f,x,\alpha')$.
One defines similarly a notion of stable function $g:D^{\bot}(\cat{S}')\pfun 
D^{\bot}(\cat{S})$, with notation $m(g,\alpha ',x)$.
\end{definition}

\begin{definition}[symmetric algorithm] \label{sym-alg-Seq}
Let \cat{S} and $\cat{S}'$ be two sds's. A symmetric 
algorithm\index{symmetric algorithm} from \cat{S} to $\cat{S}'$ is a 
pair
$$(f:D(\cat{S})\fun D(\cat{S}'),g:D^{\bot}(\cat{S}')\pfun 
D^{\bot}(\cat{S}))$$
of a function and a partial function that are both continuous and 
satisfy the following axioms:
\begin{enumerate}
\item[(L)] $(x\in 
D(\cat{S}),\alpha'\in\cpt{D^{\bot}(\cat{S}')},\wins{f(x)}{\alpha'}) 
\Rightarrow \left\{\begin{array}{l} \wins{x}{g(\alpha ')}\mand \\
 m(f,x,\alpha')=\plays{x}{g(\alpha ')}
\end{array}\right.$

\item[(R)] $(\alpha'\in 
D^{\bot}(\cat{S}'),x\in\cpt{D(\cat{S})},\winc{x}{g(\alpha ')}  \Rightarrow 
\left\{\begin{array}{l}
\winc{f(x)}{\alpha '}\mand \\
m(g,\alpha ',x)=\playc{f(x)}{\alpha'}
\end{array}\right.$
\end{enumerate}
\medskip
\noindent
We set as a convention, for any $x$ and any $\alpha'$ such that 
$g(\alpha')$ is undefined:
$$\wins{x}{g(\alpha')}\mand \plays{x}{g(\alpha')}=\emptyset.$$
Hence the conclusion of $(L)$ is simply $m(f,x,\alpha')=\emptyset$ when 
$g(\alpha')$ is undefined. In 
contrast, when we write $\winc{x}{g(\alpha')}$ in $(R)$, we assume that 
$g(\alpha')$ is defined.
\end{definition}
Thus, $g$ provides the witnesses of stability of $f$, and conversely. Moreover, the above definition is powerful enough to imply other key properties of $f$ and $g$.
\begin{definition}
A (continuous) function 
$f:D(\cat{S})\fun D(\cat{S}')$ is called
{\em sequential} if, for any pair 
$(x,\alpha')\in\cpt{D(\cat{S})}\times \cpt{D^{\bot}(\cat{S}')}$
such that $\winc{f(x)}{\alpha'}$ and $\wins{f(z)}{\alpha'}$ for some 
$z\geq x$, there exists 
$\alpha\in\cpt{D^{\bot}(\cat{S})}$, called a {\em sequentiality 
index} of $f$ at 
$(x,\alpha')$, such that $\winc{x}{\alpha}$ and for any $y\geq x$, 
$\wins{f(y)}{\alpha'}$ implies 
$\wins{y}{\alpha}$.
\end{definition}
\begin{proposition} \label{LS-RS-Seq}
Let $f$ and $g$ be as in the previous definition. Then $f$ and $g$ 
are {\em affine} (i.e., preserve the least upper bounds of pairs of upper bounded elements) and  
satisfy the following two axioms:
\begin{enumerate}
\item[(LS)] If $x\in D(\cat{S})$, 
$\alpha'\in\cpt{D^{\bot}(\cat{S}')}$,$\winc{f(x)}{\alpha'}$, and 
$\wins{f(y)}{\alpha'}$ for some $y>x$, then $\winc{x}{g(\alpha')}$, 
and $\playc{x}{g(\alpha')}$ 
is a sequentiality index of $f$ at $(x,\alpha')$.

\item[(RS)] If $\alpha'\in D^{\bot}(\cat{S}')$, 
$x\in\cpt{D(\cat{S})}$, $\wins{x}{g(\alpha')}$, and 
$\winc{x}{g(\beta')}$ for some $\beta'>\alpha'$, then 
$\wins{f(x)}{\alpha'}$, and 
$\plays{f(x)}{\alpha'}$ is a sequentiality index of $g$ at 
$(\alpha',x)$. Hence $f$ and $g$ are sequential, and $g$ provides the witnesses of sequentiality for $f$ and conversely.
\end{enumerate}
\end{proposition}

\medskip
We turn to the composition of affine algorithms.
\begin{definition} \label{sym-comp-def-Seq}
Let \cat{S}, $\cat{S}'$ and $\cat{S}''$ be sds's, and let $(f,g)$ and 
$(f',g')$ be symmetric 
algorithms from \cat{S} to $\cat{S}'$ and from $\cat{S}'$ to 
$\cat{S}''$. We define their 
composition $(f'',g'')$ from $\cat{S}$ to $\cat{S}''$ as follows:
\begin{eqnarray*}
f''=f'\comp f &\mand\; &g''=g\comp g'.
\end{eqnarray*}
\end{definition}

The announced full abstraction theorem is the following.
\begin{theorem} \label{sym-strat-Seq}
The sets of affine algorithms and of symmetric algorithms are in a bijective correspondence (actually, an isomorphism), and the two definitions of composition coincide up to the correspondence.
\end{theorem}

We just briefly indicate how to pass from one point of view to the other. Given $\phi\in D(\cat{S}\limpl\cat{S}')$, we define a pair $(f,g)$ of a 
function and a partial function as follows:
$$\begin{array}{lll}
f(x) & = & \setc{r'}{r'=s\rst_{\cat{S}'}\mand s\rst_\cat{S}\in 
x\mbox{ for some }s\in\phi}\\
g(\alpha') & = & \setc{q}{q=s\rst_\cat{S}\mand 
s\rst_{\cat{S}'}\in\alpha'\mbox{ for some }s\in\phi}.
\end{array}$$
(By convention, if the right hand side of the 
definition of $g$ is empty for some $\alpha'$, we interpret 
this definitional equality as saying that $g(\alpha')$ is undefined.)

Conversely, given a symmetric algorithm $(f,g)$ from \cat{S} to 
$\cat{S}'$, we construct an affine algorithm $\phi\in 
D(\cat{S}\limpl\cat{S}')$ by building the 
positions $s$ of $\phi$ by induction on the length of $s$ (a streamlike process!).
For example, if $s\in\phi$, if $s\rst_\cat{S}$ and $s\rst_{\cat{S}'}$ are 
responses, and if $q'=(s\rst_{\cat{S}'})c'$ 
for some $c'$, then:
$$\begin{array}{ll}
sc'c\in\phi & \mif (s\rst_\cat{S})c\in g(q')\\
sc'v'\in\phi & \mif q'v'\in f(s\rst_\cat{S})\; .
\end{array}$$

\medskip
But, as remarked above, we do not get all sequential functions in this way. Recall that in linear logic the usual implication $A\Rightarrow B$ is decomposed as
$(!A)\limpl B$  (!, and its de Morgan dual ?, are called {\em exponentials} in linear logic).

\begin{definition}[exponential] \label{exponential-def-Seq}
Let $\cat{S}=(C,V,P)$ be an sds.  We set $!\cat{S}=(Q,R,P_!)$, where $Q$ and $R$ are the sets of queries and of responses of $\cat{S}$, respectively, and where $P_!$ is recursively specified as follows (letting $\rho$ range over responses in $P_!$):
$$\begin{array}{ll}
\rho q\in P_! & \mif q\in A({\tt strategy}(\rho))\\
\rho q(qv)\in P_! & 
\mif \rho q\in P_!, {\tt strategy}(\rho q(qv))\in D(M), \mand qv\nin {\tt strategy}(\rho)
\end{array}$$
where ${\tt strategy}$ is the following function mapping responses (or 
$\epsilon$) of $P_!$ to strategies of $\cat{S}$:
$$\begin{array}{lll}
{\tt strategy}(\epsilon) =\emptyset & &
{\tt strategy}(\rho q(qv))={\tt strategy}(r)\union\set{qv}.
\end{array}$$
{\em Sequential algorithms} between two sds's $\cat{S}$ and $\cat{S'}$ are by definition affine algorithms between $!\cat{S}$ and $\cat{S'}$. 
\end{definition}

It is easily checked that the programs for $lor$ (cf. example \ref{affine-algo-ex-Seq}), $ror$, ${\it lsor}$, and ${\it rsor}$ transcribe as sequential algorithms from $Bool\times Bool$ to $Bool$.

 Sequential algorithms also enjoy two direct definitions, a concrete one and an abstract one, and both an operational and a denotational definition of composition, for which full abstraction holds, see
\cite{Cur86}.

\medskip
Let us end the section with a criticism of the terminology of symmetric algorithm. As already pointed out, the pairs $(f,g)$ are not quite symmetric since $g$ unlike $f$ is a {\em partial} function.
Logically, $\cat{S}\limpl\cat{S'}$ should read as $\cat{S}^\bot\lpar\cat{S'}$. But something odd is going on: the connective $\lpar$ would have two arguments of a different {\em polarity}: in $\cat{S'}$ it is Opponent who starts, while Player starts in $\cat{S}^\bot$. 
For this reason, Laurent proposed to decompose the affine arrow \cite{Laurent2002} (see also \cite{Boudes}):
$$\cat{S}\limpl\cat{S'}=(\downarrow\cat{S})^\bot\lpar\cat{S'}$$ 
where $\downarrow$ is a change of polarity operator. For sds's, this operation is easy to define: add a new initial opponent move, call it $\star$, and prefix it to all the positions of $\cat{S}^\bot$. For example,
$\downarrow (Bool^\bot)$ has  $\star\:?\:{\it tt}$ and  $\star\:?\:{\it ff}$ as (maximal) positions.
According to Laurent's definition, the initial moves of $\cat{S_1}\lpar\cat{S_2}$ are pairs $(c_1,c_2)$ of initial (Opponent's) moves of   $\cat{S_1}$ and $\cat{S_2}$. Then the positions continue as interleavings of a position of $\cat{S_1}$ and of $\cat{S_2}$. Notice that this is now completely symmetric in $\cat{S_1}$ and $\cat{S_2}$. 

Now, let us revisit the definition of $\cat{S}\limpl\cat{S'}$. We said that the positions of this sds had to start with a $c'$, which is quite dissymetric. But the $\downarrow$ construction allows us to 
restore equal status to the two components of the $\lpar$. A position in $\cat{S}^\bot\lpar\cat{S'}$ must start with two moves played together in $\cat{S}$ and $\cat{S'}$. It happens that these moves have necessarily the form $(\star,c')$, which is conveying the same information as $c'$.

\section{Related works} \label{rel-work}

Sequential algorithms turned out to be quite central in the study of sequentiality. 
First, let us mention that Kleene has developed (for lower types) similar notions \cite{Kleene}, under the nice name of oracles, in his late works on the semantics of 
higher order 
recursion theory (see \cite{Buccia} for a detailed comparison).

Two important models of functions that have been constructed since turned out to be the extensional collapse (i.e. the hereditary quotient equating sequential algorithms computing the same function, i.e. (in the affine case) two algorithms $(f,g)$ and $(f',g')$ such that $f=f'$): Bucciarelli and Ehrhard's model of strongly stable functions \cite{BuEhrSS,EhrCol}, and Longley's model of sequentially realizable functionals \cite{Longley}. The first model arose from an algebraic characterization of sequential (first-order) functions, that carries over to all types. The second one is a realizability model over a  combinatory algebra in which the interaction at work in sequential algorithms is encoded.

Also, Laird has shown that sequential algorithms can be obtained by a collapsing construction from his games model of control in Hyland and Ong style \cite{Laird}. 

Hyland and Ong's model and Abramsky-Jagadeesan-Malacaria's model  (HO and AJM, respectively) capture $\pcf$ definability exactly, whereas the games
 associated with sequential algorithms also accommodate 
control operations such as {\tt call-cc} that are not definable in $\pcf$ (see section \ref{control}). In fact, the interpretation function from normal forms to these models is injective.  An essential merit of these works was to characterize the image of this injection, and hence to characterize $\pcf$ definability in a syntax-independent way by a few conditions such as innocence and well-bracketing. This opened the way to a whole research program launched by Abramsky. What does happen if one of the conditions is relaxed? Giving up innocence led to very interesting (and fully abstract) models of  references (see \cite{AMC}). Giving up well-bracketing gave a model of $\pcf$ plus control, as already mentioned.

\medskip
The model of sequential algorithms and the HO (or AJM) model
differ drastically in size. The type $Bool\fun Bool$ is interpreted 
by a finite sds (i.e., an sds with finitely many positions) in the model 
of sequential algorithms, while there are infinitely many $\pcf$ B\"ohm 
trees (and hence infinitely many strategies) in the HO and AJM models
at that type.  The difference comes from the way the exponential is defined.
In definition \ref{exponential-def-Seq}, a key feature is non-repetition ($ qv\nin {\tt strategy}(\rho)$).
In the games models, the exponential is defined either by interleaving allowing for repetitions or by the opening of potentially infinitely many copies of positions. Roughly, this amounts to dropping the condition $qv\nin {\tt strategy}(\rho)$.

The finitary nature of sequential algorithms implies that 
equality
in the model is
decidable for any type built over $Bool$, 
while the term model games do not provide effective tools
to tackle observational equivalences. As a matter of fact,
 it has been proved by Loader \cite{Loa} that equality of two objects in the fully  abstract model of (finitary) $\pcf$ is
undecidable.  A model of $\pcf$ is called fully abstract  if it equates two terms if and only if these terms are {\em observationally equivalent}, which means that one can be replaced by the other in any program context without affecting the final result.  The {\em full abstraction problem} of Scott, Milner and Plotkin was the (quite loosely) specified problem of providing a {\em denotational}  construction of the fully abstract model of \pcf, as opposed to the original term-based  construction of Milner, who also had shown the uniqueness of the fully abstract model \cite{Mil77}. The HO and AJM games models  can be called denotational, since they provide a syntax-independent characterization of a term model made of 
(a $\pcf$ version of) B\"ohm trees. But they yield full abstraction only via a collapse construction which is not essentially different from the one originally performed by Milner. An implicit hope was to arrive at decidability results for the equality in the model, as  usual denotational models  consist of functions, and hence interpret every type built over Bool by a finite set. Loader's result says that there cannot be such a construction of the fully abstract model of \pcf, and justifies a posteriori why
game models had to use infinitary exponentials. In contrast, when \pcf~is extended with control, then the finitary exponential of the model of sequential algorithms does the job (coming next).

\section{Control} \label{control}

We already pointed out that theorem \ref{sym-strat-Seq} is a full abstraction result
 (for the affine case), and that  the same theorem has been proved for all sequential algorithms with respect to the language CDS. Sequential algorithms allow inherently to consult the internal behaviour of their arguments and to make decisions according to that behaviour. For example,
there exists a sequential algorithm of type $(Bool^2\rightarrow Bool)\rightarrow Bool$ that maps ${\it lsor}$ to ${\it tt}$ and ${\it rsor}$ to ${\it ff}$ (cf. end of section 2).
Cartwright and Felleisen made the connection with more standard control operators explicit, and this lead to the 
full abstraction result of sequential algorithms with respect to an extension of $\pcf$ with a control operator \cite{CCF94}. 

\medskip
In this respect, we would like to highlight a key observation made by Laird. Let $o$ be the sds
with one cell and no value: $o=\set{\set{?},\emptyset,\set{?}}$. Then we have the isomorphism
$$Bool\sim (o\rightarrow o\rightarrow o)$$
where $Bool$ is the sds $\set{\set{?},\set{{\it tt},{\it ff}},\set{?,(?\:{\it tt}),(?\:{\it ff})}}$ considered above.
Indeed, both sds's have exactly three strategies, ordered in the same way:
$$\begin{array}{l}
D(Bool)=\set{\emptyset,\set{?\:{\it tt}},\set{?\:{\it ff}}}\\ 
D(o\rightarrow o\rightarrow o)=\set{\emptyset,\set{?_\epsilon\:?_1},\set{?_\epsilon\:?_2}}
\end{array}$$
(we use subscripts to decorate the cells of the three copies of $o$, using the convention $o_1\rightarrow o_2\rightarrow o_\epsilon$). It is an instructive exercise to write down explicitly the inverse isomorphisms as sequential algorithms: in one direction, one has the ${\tt if\:then\:else}$ function, in the other direction, we have the control operation 
${\tt catch}$ considered in  \cite{CCF94}, which tells apart the two strategies 
$,\set{?_\epsilon\:?_1},\set{?_\epsilon\:?_2}$.  Here, we shall  show (at type $bool$) how the control operator {\tt call-cc} of Scheme or Standard ML is interpreted as a sequential algorithm of type
$((bool\rightarrow B)\rightarrow bool)\rightarrow bool$. The formula $((A\rightarrow B) \rightarrow A)\rightarrow A$ is called Pierce's law and is a typical tautology of classical logic. The connection between control operators and classical logic -- and in particular the fact that {\tt call-cc} corresponds to Pierce's law--  was first discovered in \cite{Grif}. Here is is the sequential algorithm interpreting {\tt call-cc} for 
$A=bool$:
$$?_\epsilon\:?_1\left\{\begin{array}{l}
?_{11}\:?_{111}\left\{\begin{array}{l}
{\it tt}_{111}\:{\it tt}_\epsilon\\
{\it ff}_{111}\:{\it ff}_\epsilon
\end{array}\right.\\
{\it tt}_1\:{\it tt}_\epsilon\\
{\it ff}_1\:{\it ff}_\epsilon
\end{array}\right.$$
(with  labelling of moves
$((bool_{111}\rightarrow B_{11})\rightarrow bool_1)\rightarrow bool_\epsilon$).
The same algorithm, with $bool$ replaced by $o\rightarrow o\rightarrow o$, is:
$$?_\epsilon\:?_1\left\{\begin{array}{l}
?_{11}\:?_{111}\left\{\begin{array}{l}
?_{1111}\:?_2\\
?_{1112}\:?_3
\end{array}\right.\\
?_{12}\:?_2\\
?_{13}\:?_3
\end{array}\right.$$
(with  labelling
$(((o_{1111}\rightarrow o_{1112})\rightarrow o_{111}\rightarrow B_{11})\rightarrow o_{12}\rightarrow_{13}\rightarrow o_1)\rightarrow o_2\rightarrow o_3\rightarrow o_\epsilon$).
The reader familiar with continuations may want to compare this tree with the continuation-passing (CPS) style interpretation $\lbd yk.y(\lbd xk'.xk)k$ of {\tt call-cc}, or in tree form (cf. section \ref{prologue}):
$$\lbd yk.y\left\{\begin{array}{l}
\lbd xk'.x\left\{\begin{array}{l}
k
\end{array}\right.\\
k
\end{array}\right.$$
where the first $k$ indicates a copy-cat from $o_{111}$ to $o_\epsilon$ while the second one indicates a copycat from $o_1$ to $o_\epsilon$. The bound variable $k'$ amounts to the fact $B$ itself is of the form $B'\rightarrow o$ (see below).
This is an instance of the injection from terms to strategies mentioned in section \ref{rel-work} (in this simple example, Laird's HO style model coincides with that of sequential algorithms).

CPS translations are the usual indirect way to interpret control operators: first translate, then interpret in your favorite cartesian closed category. In contrast, sequential algorithms look as a direct semantics. The example above suggests that this is an
``illusion'': once we explicitly replace $bool$ by $o\rightarrow o\rightarrow o$, we find the indirect way underneath. 

\medskip
A more mathematical way to stress this is through Hofmann-Streicher's notion of continuation model \cite{HofStrei}: given a category having all the function spaces $A\rightarrow R$ for some fixed object $R$ called object of final results, one only retains the full subcategory of {\em negative} objects, that is, objects of the form $A\rightarrow R$. In this category, control can be interpreted. (For the logically inclined reader, notice that thinking of $R$ as the formula ``false'',  then the double negation of $A$ reads as $(A\rightarrow R)\rightarrow R$, and the classical tautology $((A\rightarrow R)\rightarrow R)\rightarrow A$ is intuitionistically provable for all negative $A=B\rightarrow R$.)
Now, taking $R=o$, the above isomorphism exhibits $bool$ as a negative object. But then all types are negative: given $A$ and $B=B'\rightarrow R$, then
$A\rightarrow  B \sim  (A \times B')\rightarrow R$ is also negative. Hence the model of sequential algorithms (and Laird's model of control) are indeed continuation models, but it is not written on their face.

\section{A few more remarks}

We would like to mention that this whole line of research on sequential interaction induced such side effects as the design of the Categorical Abstract Machine \cite{CAM}, that gave its name to the language CAML, and of a theory of Abstract B\"ohm Trees, alluded to in section \ref{prologue}. 

\medskip
As for future lines of research, imports from and into the  program of ludics newly proposed by Girard \cite{Gir01} are expected.  We just quote one connection with ludics.
We insisted in section 
\ref{intro} that ${\it lsor}$ and ${\it rsor}$ were different programs for the same function. But there is a
way to make them into two different functions, by means of additional {\em error} values, and accordingly of additional constants in the syntax. Actually, one error is enough, call it $err$. Indeed, we have:
$${\it lsor}(err,\bot)=err \quad\quad {\it rsor}(err,\bot)=\bot\;.$$
Because ${\it lsor}$ looks at its left argument first, if an error is fed in that argument, it is propagated, whence the result $err$. Because ${\it rsor}$ looks at its right argument first, if no value is
is fed for that argument, then the whole computation is waiting, whence the result $\bot$.
One could achieve the same more symmetrically with two different errors:
${\it lsor}(err_1,err_2)=err_1$, ${\it rsor}(err_1,err_2)=err_2$.
But the economy of having just one error is conceptually important, all the more because in view of the isomorphism of section \ref{control}, we see that we can dispense (at least for $bool$ but also for any finite base type) with the basic values ${\it tt}, {\it ff}, 0, 1,\ldots$. We arrive then at a picture with only two (base type) constants: $\bot$ and $err$! This is the point of view adopted in Girard's ludics. In ludics, the counterpart of $err$ is called Daimon. The motivation for introducing Daimon is quite parallel to that of having errors. Girard's program has the ambition of giving an interactive account of proofs.
So, in order to explore a proof of a proposition $A$, one should play it against a ``proof'' of $A^\bot$ (the negation of linear logic). But it can't be a proof, since not both $A$ and $A^\bot$ can be proved.
So, the space of  ``proofs'' must be enlarged to allow for more opponents to interact with.
Similarly, above, we motivated errors by the remark that, once introduced, they allow more observations to be made: here, they allowed  us to separate ${\it lsor}$ and ${\it rsor}$. More information, also of a survey kind, can be found in \cite{sym-inter}.

\end{document}